\pdfoutput=1
\documentclass[usenatbib]{mn2e}
\usepackage[]{natbib}
\usepackage{graphicx}
\usepackage{lscape}
\usepackage{amssymb}
\usepackage{amsmath}
\usepackage{url}
\usepackage{aas_macros}
\usepackage{multirow}
\usepackage{float}
\usepackage{color}

\newcommand{\PHDLR}[1]{}
\newcommand{\ro}{$R_{\rm 0}$}

\newcommand\ion[2]{\mbox{#1\hspace{0.2em}{\rmfamily\@Roman{#2}}}}
\newcommand{\hii}{\mbox{H\,{\sc ii}}}
\newcommand{\hi}{\mbox{H\,{\sc i}}}
\newcommand{\abs}[1]{\lvert#1\rvert}
\newcommand{{\annstatist}}{The Annals of Statistics}
\newcommand{\kms}{km\,s$^{-1}$}
\newcommand{\changefunction}[1]{\expandafter\renewcommand\csname#1\endcsname[1][]%
    {\qopname\relax o{#1}\ifx\relax##1\relax\else^{##1}\fi}}
\changefunction{sin}
\changefunction{cos}
\changefunction{tan}
\renewcommand{\arcsin}{\sin[-1]}
\renewcommand{\arctan}{\tan[-1]}

\title[North Galactic Pole]{Revised Geometric Estimates of the North
  Galactic Pole and the Sun's Height Above the Galactic Midplane}

\author[]{M. T. Karim $^1$\thanks{Email:
    mkarim2@u.rochester.edu} and Eric E. Mamajek$^{1,2}$ \\
$^{1}$Department of Physics \& Astronomy, University of Rochester,
  Rochester, NY 14627, USA\\
$^{2}$Current Address: Jet Propulsion Laboratory, California Institute of Technology,
  4800 Oak Grove Dr., Pasadena, CA 91109, USA}

\begin{document}

\date{Accepted 2016 October 24. Received 2016 October 22; in original form 2016 September 5}

\pagerange{\pageref{firstpage}--\pageref{lastpage}} \pubyear{2016}

\maketitle

\label{firstpage}

\begin{abstract}
Astronomers are entering an era of $\mu$as-level astrometry utilizing
the 5-decade-old IAU Galactic coordinate system that was only
originally defined to $\sim$0$^{\circ}$.1 accuracy, and where the
dynamical centre of the Galaxy (Sgr A*) is located
$\sim$0$^{\circ}$.07 from the origin.
We calculate new independent estimates of the North Galactic Pole (NGP)
using recent catalogues of Galactic disc tracer objects such as embedded 
and open clusters, infrared bubbles, dark clouds, and young massive stars.
Using these catalogues, we provide two new estimates of the NGP.
Solution 1 is an ``unconstrained'' NGP determined by the galactic
tracer sources, which does not take into account the location of Sgr
A*, and which lies $90^{\circ}.120\,\pm\,0^{\circ}.029$ from Sgr A*,
and Solution 2 is a ``constrained'' NGP which lies exactly
90$^{\circ}$ from Sgr A*.
The ``unconstrained'' NGP has ICRS position:
$\alpha_{NGP}$ = $192^{\circ}.729$ $\,\pm\,$ $0^{\circ}.035$, 
$\delta_{NGP}$ = $27^{\circ}.084$ $\,\pm\,$ $0^{\circ}.023$ and 
$\theta\,$ = $122^{\circ}.928$ $\,\pm\,$ $0^{\circ}.016$.
The ``constrained'' NGP which lies exactly 90$^{\circ}$
away from Sgr A* has ICRS position:
$\alpha_{NGP}$ = $192^{\circ}.728 \,\pm\, 0^{\circ}.010$, 
$\delta_{NGP}$ = $26^{\circ}.863 \,\pm\, 0^{\circ}.019$ and 
$\theta\,$ = $122^{\circ}.928$ $\,\pm\,$ $0^{\circ}.016$.
The difference between the solutions is likely due to the Sun lying
above the Galactic midplane.
Considering the angular separation between Sgr A* and our
unconstrained NGP, and if one adopts the recent estimate of the
Galactocentric distance for the Sun of \ro\, = 8.2\,$\pm$\,0.1 kpc,
then we estimate that the Sun lies $z_{\odot}$ $\simeq$ 17\,$\pm$\,5
pc above the Galactic midplane.
Our value of $z_{\odot}$ is consistent with the true median of 55
previous estimates published over the past century of the Sun's height
above the Galactic mid-plane ($z_{\odot}$ $\simeq$ 17\,$\pm$\,2 pc).
\end{abstract}

\begin{keywords}
  The Galaxy (centre, disc, fundamental parameters, general,
  kinematics and dynamics, structure)
\end{keywords}


\section{Introduction}
\label{Introduction}

{\it ``The anticipated improvement in the position of the [Galactic] pole, by a
  factor of 10/1 or better, is substantial, and is unlikely to be
  increased for many years''} \citep{Blaauw60}\\

While the need for a Galactic Coordinate System (GCS) goes back at
least to the time of William Herschel \citep{Herschel1785}, the first
standard Galactic coordinate system in common international use was
that of \citet{Ohlsson32}.
The GCS is defined as a spherical celestial coordinate system with
right ascension ($\alpha_{P}$) and declination ($\delta_{P}$) marking
the North Galactic Pole (NGP), and its equator tracing the Galactic
plane.
In the early 20th century, the zero longitude ($\ell$ = 0$^{\circ}$)
was defined between this pole and the point where the galactic plane
intercepted the equinox 1900.0 celestial equator \citep{Ohlsson32,
vanTulder42,Ohlsson56}.
\citet{Ohlsson32} determined the location of the NGP to be
$\alpha_{P}$ = 12$^h$40$^m$, $\delta_{P}$ =
+28$^{\circ}$ (B1900.0).
\citet{vanTulder42} later determined a NGP using numerous samples of
optical stars, tracing extinction in the Galactic plane, and other
indicators, which deviated from Ohlsson's pole by $\sim$1$^{\circ}$
$\alpha_{P}$ = 12$^h$44$^m$ ($\pm$0$^{\circ}$.3), $\delta_{P}$ =
+27$^{\circ}$.5 ($\pm$0$^{\circ}$.2) (B1900.0 equinox).
By the mid-1950's, it was apparent from \hi\, surveys and the
discovery of an obvious dynamical centre to the Galaxy (associated
with the radio source Sgr A) that a revision to the
\citet{vanTulder42} GCS was needed.\\


At the Xth IAU General Assembly in Moscow, IAU Commissions 33 (Stellar
Statisics) and 40 (Radio Astronomy) authorized the formation of an IAU
sub-commission (33b) to determine a new GCS which principally traced
the neutral hydrogen in the Galactic
plane\footnote{https://www.iau.org/static/resolutions/IAU1958\_French.pdf}.
Summaries of the research by IAU Sub-Commission 33b defining the IAU
GCS were given in a series of contemporaneous papers by \citet[][Paper
  I]{Blaauw60}, \citet[][Paper II]{Gum60}, \citet[][Paper
  III]{Gum60B}, \citet[][Paper IV]{Blaauw60B} and \citet[][Paper
  V]{Oort60}.
Paper I \citep{Blaauw60} summarized the sub-commission's research and
comprises their proposal of the 1958 revised GCS to the IAU.
Using \hi\, survey data by groups at Leiden and Sydney, Paper II
\citep{Gum60} demonstrated the flatness of the Galactic neutral
hydrogen, determined the best-fitting \hi\, Galactic plane, and
independently estimated that the Sun was near the plane, within
uncertainties (height $z_{\odot}$ = 4\,$\pm$\,12 pc).
Further, Paper III \citep{Gum60B} argued that the trends in intensity
for \hi\, and radio continuum data were consistent with Sgr A marking
the centre of the Galaxy.
Paper IV \citep{Blaauw60B} argued that optical stars (e.g. OB stars,
Cepheids, etc.)  should not be employed to constrain the NGP due to
limited sampling and interstellar extinction, and that these samples
appear to be coplanar with the \hi\, principle plane.
Paper V \citep{Oort60} presented further arguments based on optical
and radio data for positing Sgr A to mark the dynamical centre of the
Galaxy.
Based on Blaauw's (1960) argument that the \hi\, plane was a superior
means of defining the Galactic plane and NGP (abandoning van Tulder's
optical-based methodology), and employing the \hi\, plane defined by
\citet{Gum60} to define the Galactic plane, and defining the radio
source Sgr A as the logical choice of origin \citep{Gum60B, Oort60},
the commission proposed the IAU GCS to have NGP at $\alpha^{IAU}_{P}$
= 12$^h$49$^m$, $\delta^{IAU}_{P}$ = +27$^{\circ}$.4 with position
angle $\theta^{IAU}_{P}$ = 123$^{\circ}$ \citep[all
  B1950;][]{Blaauw60}.
The position angle $\theta$ is that of the {\it ``the new galactic
  pole of the great circle passing through Sagittarius A''}
\citep{Blaauw60}.
On the ICRS, the IAU NGP corresponds to $\alpha^{IAU}_{P}$ =
12$^h$51$^m$26$^s$.2755, $\delta^{IAU}_{P}$ = +27$^{\circ}$07'41''.704
and $\theta^{IAU}_{P}$ = 122$^{\circ}$.93191857 in J2000.0
\citep{Liu11ir}.
Furthermore, the position of the Galactic Centre (GC) on the IAU GCS
was set to be at $\alpha^{IAU}_{GC}$ = 17$^h$42$^m$37$^s$,
$\delta^{IAU}_{GC}$ = -28$^{\circ}$57' \citep[B1950;][]{Gum60}.
On the J2000 system this position corresponds to $\alpha^{IAU}_{GC}$ =
17$^h$45$^m$37$^s$.224, $\delta^{IAU}_{GC}$ = -28$^{\circ}$56'10".23
\citep{Reid04}.\\


However, radio surveys in recent decades have been steadily improving
our understanding of Sgr A, resolving the bright nonthermal radio
source Sgr A* as the true dynamical centre of the Galaxy
\citep[e.g.][]{Brown82,Reid04}.
\citet{Reid04} determined the precise position of Sgr A* to be
$\alpha_{GC}$ = 17$^h$45$^m$40$^s$.0409 , $\delta_{GC}$ =
-29$^{\circ}$00'28'.118 (ICRS, epoch 2000.0) with proper motion
$\mu_{\alpha}$ = -3.151\,$\pm$\,0.018 mas\,yr$^{-1}$ and
$\mu_{\delta}$ = -5.547\,$\pm$\,0.026 mas\,yr$^{-1}$.
Hence, there is an offset of 0$^{\circ}$.0724 when comparing the
position of the origin of the IAU GCS and Sgr A*.
Given the promised improvement in astrometric precision afforded by
Gaia \citep{Lindegren16}, and the ability of the survey catalogue to
greatly improve our knowledge of Galactic dynamical parameters
\citep{Bland-Hawthorn16}, one would prefer that Galactic positions
and velocities could be calculated accurately on a natural Galactic
coordinate system to at least third significant figure.
Hence, it can be argued that the IAU 1958 GCS is no longer an adequate
representation of the Milky Way's natural orientation.\\


There has been little published work on refining the Galactic
coordinate system since the 1950s IAU effort.
Recently \citet{Liu2011} tried to determine the transformation matrix in the framework
of ICRS but could not find a GCS that is connected steadily to the ICRS.
Later \citet{Liu11ir} determined estimates of the NGP based on the
Two-Micron All-Sky Survey (2MASS) and SPECFIND v2.0 catalogues.
More recently \citet{Ding15} solved for new estimates of the NGP
based on points sources in two infrared sky surveys from the AKARI and
WISE missions (covering bandpasses between 3.4\,$\mu$m and
90\,$\mu$m) using the same method as \citet{Liu11ir}.\\


In this paper, we provide independent estimates of the North Galactic
Pole using less biased tracers of Galactic structure than have been
often used before (e.g. \hi, near-IR star counts, etc.).
Section \ref{sec:Data} discusses the classes of Galactic tracers that
we employ and the catalogues of tracers selected for analysis.
Section~\ref{sec:Analysis} discusses how we calculated the fundamental
parameters, i.e. the RA and the Dec of the NGP and the position angle
of the NGP with respect to the Galactic Centre at the North Celestial
Pole. 
Lastly, Section~\ref{sec:Discussion} compares our estimates of these
Galactic parameters to previously published studies, and presents an
estimate of the Sun's height above the Galactic midplane based on our
revised NGP.

\section{Data}
\label{sec:Data}

\subsection{Categories of Galactic Tracers}
\label{sec:Categories}

Throughout this study, we consider ``{\it Galactic tracers}" as
classes of astronomical objects that strongly trace the Galactic disc
and plane, with obvious examples being \hii\, regions, embedded
clusters, infrared dark clouds (IRDCs), etc.
The current IAU definition \citep{Blaauw60} depends heavily on the
distribution of \hi\, gas - which has the advantage of tracing mass at
large distances, and demonstrates a high degree of flatness in the
plane.
Instead of relying too heavily on one tracer, we analyse multiple
classes of Galactic tracers whose samples are dominated by objects
discovered in recent decades mostly via infrared and radio surveys.
Here we discuss the different classes of Galactic tracers.\\


{\it Infrared Dark Clouds, Infrared Embedded Sources, Young Stellar
  Objects, Embedded Clusters, Open Clusters:} Stars form in embedded
clusters within molecular cloud complexes, which are typically within
a few hundred pc of the Galactic plane. Embedded clusters disperse
their dense gas on time-scale of $\lesssim$10$^{6.5}$ yr, and the
majority dissolve into unbound stellar associations on time-scale of
$\sim$10$^7$ yr (e.g. the OB associations), while $\sim$10\%\, survive
on time-scale of $\sim$10$^{8-9}$ yr as open clusters
\citep{Lada03}. All of these phenomena (molecular clouds, embedded
clusters, young stellar objects, open clusters) generally trace the
Galactic plane on larger scales, however local variations are obvious
on smaller scales (e.g. the Gould Belt).\\


A particular class of molecular cloud can be traced especially
strongly along the Galactic plane. IRDCs were discovered in the mid
1990s with the Infrared Space Observatory (ISO) and Midcourse Space
Experiment \citep[MSX;][]{Egan98}. They are observed in silhouette
against the bright infrared emission of the Galactic plane and the
most likely distance range to these clouds is 2-8 kpc
\citep{Egan98}. These clouds are cold ($<$ 25 K) and known to be the
sites of the earliest stages of star formation \citep{Frieswijk10}. We
chose IRDCs as one of the tracers because of the abundance of IRDCs in
the Galactic plane, particularly near the Galactic Centre
\citep{Chambers10}.\\


{\it \hii\, Regions, \hi\, Shells, Infrared Bubbles:} Different
flavours of ``bubbles'' pervade the interstellar medium in disc of the
Galaxy, and can be found via observations at a wide range of
wavelengths, especially the radio and infrared
\citep[e.g.][]{Anderson14, Simpson12, Churchwell06}.  \hii\, regions
are expanding high pressure regions of expanding ionized gas
associated with luminous stars outputting copious amounts of UV
radiation. Given their short dynamical time-scale and association with
short-lived massive stars, they trace star formation and are
concentrated in the spiral arms. Their strong radio emission enables
the tracing of young massive stars at large distances
\citep{Paladini03}. Bubbles detected at infrared wavelengths mostly
trace polycyclic aromatic hydrocarbon (PAH) emission, with cavities
evacuated of dust.  Only about a quarter are coincident with \hii\,
regions, and the majority appear to be dominated by mid to late B-type
stars which do not excite \hii\, regions \citep{Churchwell06}.\\


{\it AGB stars:} Asymptotic giant branch (AGB) stars are luminous
evolved stars undergoing extreme mass loss, and exhibiting very red
infrared colors due to circumstellar dust.  The AGB stars are
categorized as ``standard'' O-rich and C-rich, and ``extreme''
\citep{Robitaille08}. Such red giant stars trace the radial and
vertical structure of the Galactic disc
\citep{Benjamin05,Robitaille08}.\\


{\it Supernovae Remnants:} Supernova remnants (SNRs) are typically
detected in optical, radio, and X-ray surveys as expanding regions of
hot, shocked plasma associated with stellar explosions \citep{Green09,
  Green14}.  They are morphologically classified as ``shell'' (S),
``filled-centre'' (F), or ``composite'' (C) structures in the radio.
SNRs trace the Galactic plane fairly well in the 4th and 1st Galactic
quadrants \citep[see Fig. 3 of ][]{Green09}, however there are only
about 294 Galactic SNRs known \citep{Green14}, hence they provide a
much smaller sample of tracers compared to our other categories of
Galactic tracers.\\

\subsection{Catalogues}

We used three criteria to select our catalogues for analysis:
\begin{itemize}
  
\item They must be composed of the tracers mentioned in
  Section~\ref{sec:Categories}.

\item The catalogues should map the Galactic plane reasonably evenly.

\item For a given tracer, the corresponding catalogue should be the
  most recent and comprehensive one.

\end{itemize}
Coordinates for the tracers drawn from the catalogues are on the
International Celestial Reference System (ICRS) for J2000.
The selected catalogues are discussed in the following paragraphs.\\
%


\citet{Anderson14} identified previously known Galactic \hii\, regions
and newly discovered candidate \hii\, region using data from the
all-sky \textit{WISE} survey \citep{Wright10}. This catalogue was created by
visually inspecting \textit{WISE} 12-$\mu$m and 22-$\mu$m images
spanning the entire Galactic plane within the galactic latitude range
$\abs{b} \leq 8^{\circ}$. This catalogue consists of 8399 \hii\,
regions (either previously known objects or new candidates).\\


\citet{Csengeri14} identified embedded sources throughout the inner
Galaxy using the \textit{ATLASGAL} survey. The \textit{ATLASGAL}
survey imaged the Galactic Plane between Galactic longitude
$-80^{\circ}<l<60^{\circ}$ and Galactic latitude $-2^{\circ}\, <\, b\,
<\, 1^{\circ}$ at 870-$\mu$m.  These embedded sources are identified
as the most prominent star-forming regions in the Galaxy. A total of
10861 compact sources were compiled in this catalogue.\\


\citet{Lumsden13} assembled a large catalogue of statistically
selected young massive protostars and \hii\, regions. It was
constructed using mid- and near-infrared data from the \textit{MSX}
and \textit{2MASS} surveys, respectively \citep{Egan98,
  Skrutskie06}. The mid-IR survey was done in $b<5^{\circ}$ and
$10^{\circ} < l < 350^{\circ}$, and the catalogue contains 2811
objects.\\


\citet{Morales13} catalogued 695 known open and embedded clusters in
the inner Galaxy region. This catalogue was created from the
\textit{ATLASGAL} survey and the identified clusters that are within
the range $\abs{l} \le 60^{\circ}$ and $\abs{b} \le 1.5^{\circ}$.
Most of the open clusters lie within $\sim$1 kpc, and most of the
embedded clusters lie within $\sim$2 kpc, hence the catalogue may
carry some bias in that it is more heavily represented by local
objects than the other catalogues.\\


\citet{Peretto09} identified and characterized a complete sample of
Infrared Dark Clouds (IRDCs) in the \textit{GLIMPSE}
\citep{Churchwell09} and \textit{MIPSGAL} \citep{Carey09} surveys
undertaken by the \textit{Spitzer Space Telescope}.  Their survey
analysed IRDCs in \textit{GLIMPSE} 8-$\mu$m and \textit{MIPSGAL}
24-$\mu$m images.  The \textit{GLIMPSE} survey\footnote{See
  http://www.astro.wisc.edu/glimpse/all\_GLIMPSE-data-AAS2013.pdf} is
a conglomeration of multiple surveys that covered the entire
360$^\circ$ longitude of the Galactic Plane and width ranging between
2$^\circ$-9$^\circ$ in latitude.  The \textit{MIPSGAL}
survey\footnote{See http://mipsgal.ipac.caltech.edu/a\_mipsgal.html}
complemented the \textit{GLIMPSE} legacy surveys and covered $\abs{b}
< 1$ for $5 < l < 63$ and $298 < l < 355$ of the Galactic Plane and
strips from $1 < \abs{b} < 3$ for $-5 < l < 7$ at 24-$\mu$m and
70-$\mu$m. The entries of the catalogue span $10^{\circ} < l <
65^{\circ}$ and $\abs{b} < 1^{\circ}$. A total of 11303 clouds were
analysed in this paper, of which 80\%\, of the IRDCs were newly
identified. \\


\citet{Robitaille08} catalogued sources in the Galactic midplane from
the \textit{GLIMPSE} survey that have intrinsically red mid-infrared
colours. These sources are dominated by high- and intermediate-mass
young stellar objects (YSOs) and asymptotic giant branch (AGB)
stars. Planetary nebulae and background galaxies together represent at
most 2-3\% of all the sources. This catalogue of 18,949 sources
represents one of the largest compilations of YSOs and AGB stars
tracing the Galactic plane.\\


\citet{Simpson12} catalogued bubbles in the \textit{Spitzer}
\textit{GLIMPSE} and \textit{MIPSGAL} surveys. Their survey of the
region $\abs{l}<65^{\circ}$ and $\abs{b}<1^{\circ}$ was based on
visual inspection based on the online citizen scientists via ``The
Milky Way Project''.  This catalogue divides the bubbles into two
categories: Large bubbles and Small bubbles. We calculated the
position of the pole for the large and the small bubbles separately,
and then for the combined samples. This catalogue consists of 5106 IR
bubbles -- 3744 large bubbles and 1362 small bubbles. \\


\citet{Green14} presented a catalogue of 294 Galactic supernovae
remnants. This catalogue is heterogeneous and contains some objects
which are probable SNR candidates.  Most of the objects were detected
via radio observation, and detection of distant SNRs is hampered by
Galactic absorption. A few SNRs from this catalogue have high Galactic
latitude value, however no attempt was made to clip the relatively
small sample.\\

\section{Analysis}
\label{sec:Analysis} 


Three fundamental parameters define the GCS in spherical coordinates
on the International Celestial Reference System (ICRS): the coordinate
position of the NGP ($\alpha$ and $\delta$), and the position angle
$\theta$ of the NGP with respect to the Galactic centre (defining the
origin of the GCS), which defines longitude $\ell$ = 0$^{\circ}$.
Following the principles of the 1958 IAU determination of the GCS, the
problem can be reduced to one of solving for the best-fitting Galactic
plane (the normal vector of which defines the NGP), and defining the
origin of the GCS (determining the Galactic centre).
We estimate these parameters using two different methods. 
The first ``unconstrained'' method calculates the parameters using
least-squares fit for defining the Galactic plane through the samples
of Galactic tracers discussed in Section~\ref{sec:Data}.
The second ``constrained'' method uses the methodology of the first,
but constrains the NGP to be exactly $90^{\circ}$ away from Sgr A* by
considering the position angle calculated in
Section~\ref{subsec:Method1}.
Both of these methods are discussed in detail in the following
subsections.\\


\begin{figure}
\includegraphics[scale=0.5]{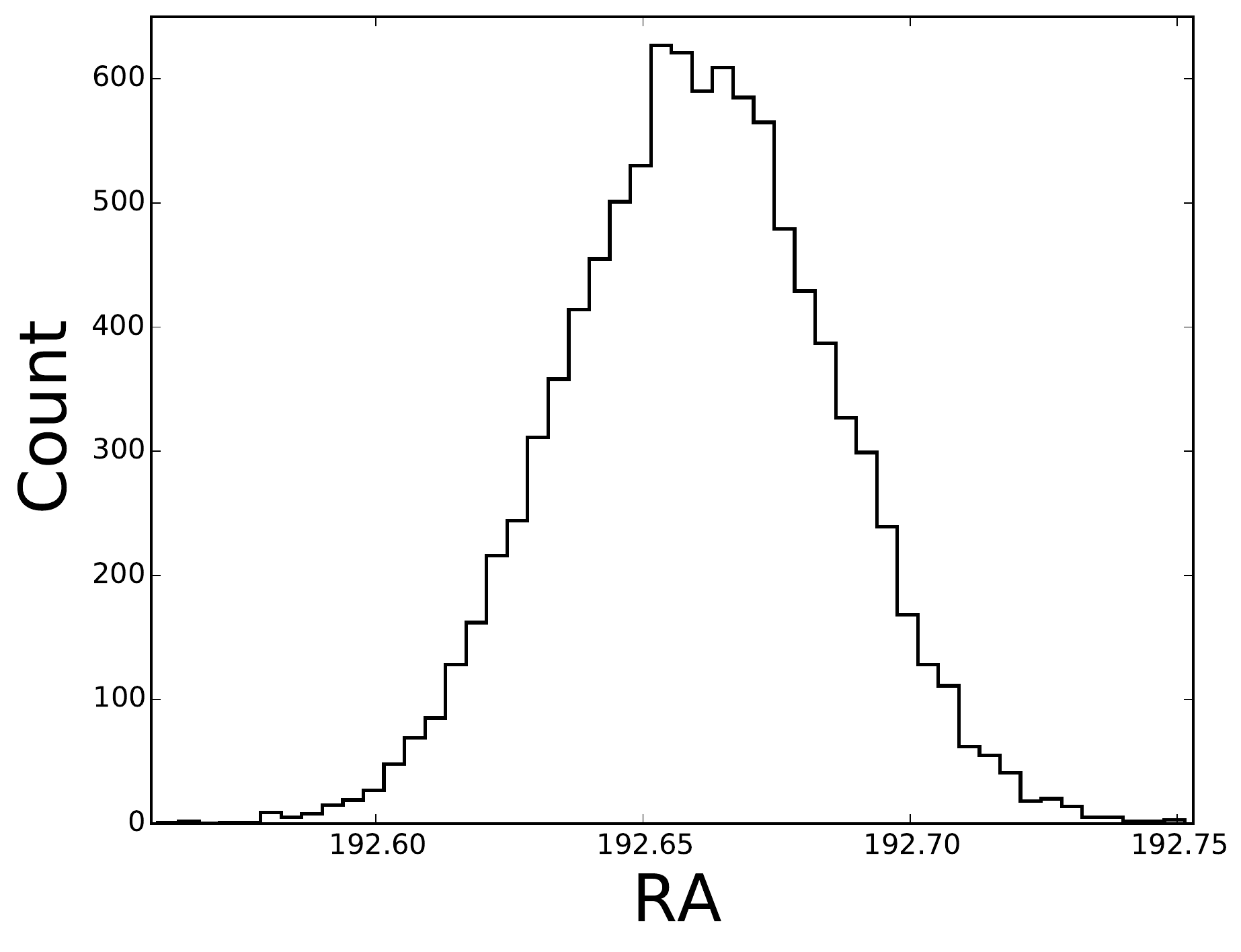}
\caption[]{Histogram of $\alpha$ for the estimated North Galactic Pole
  using for the bootstrapped \citet{Robitaille08} sample of candidate
  YSOs and AGB stars from the GLIMPSE survey.}
\label{fig:hist_rob_ra}
\end{figure}


\begin{figure}
\includegraphics[scale=0.5]{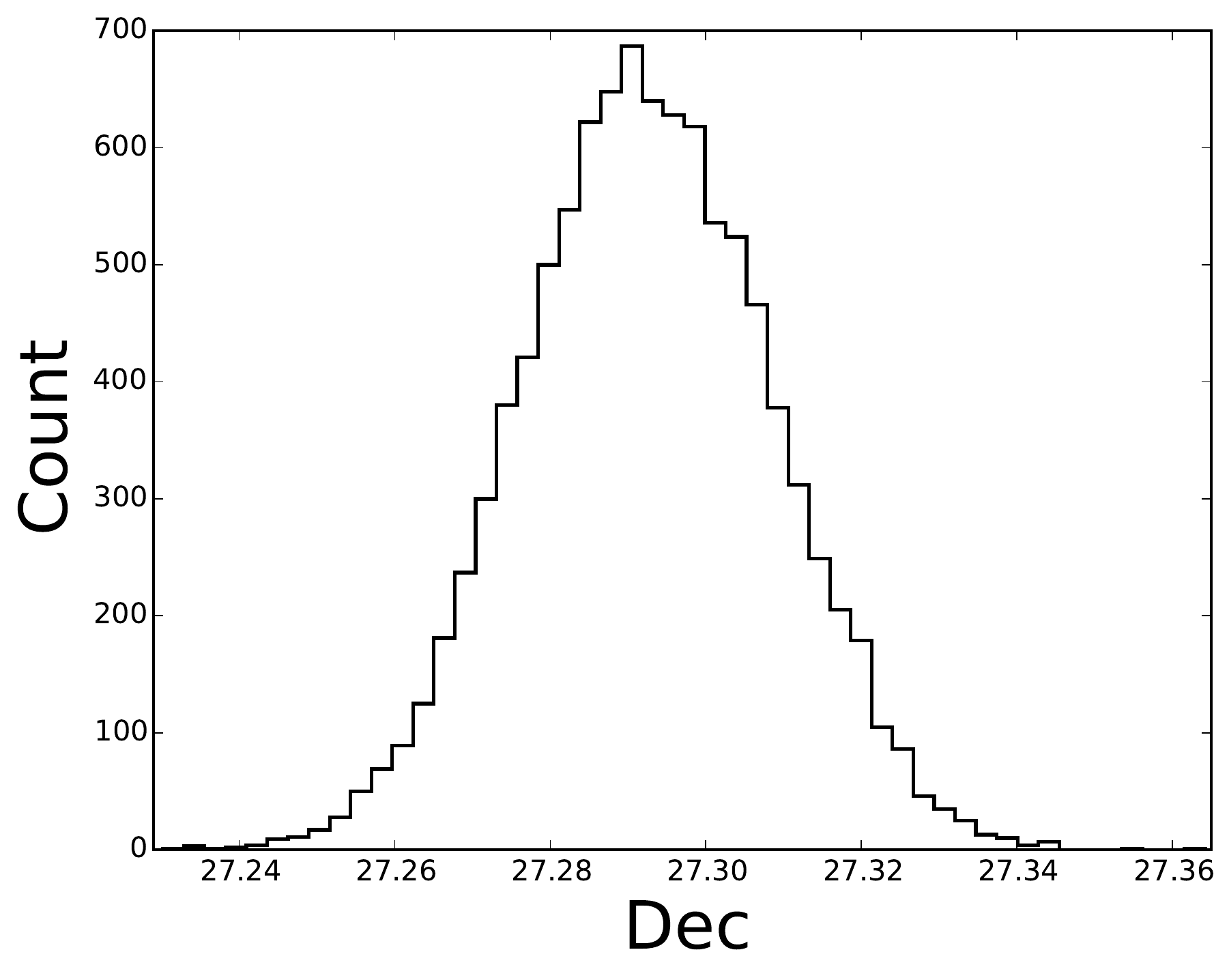}
\caption[]{Histogram of $\delta$ for the estimated North Galactic Pole
  using for the bootstrapped \citet{Robitaille08} sample of candidate
  YSOs and AGB stars from the GLIMPSE survey.}
\label{fig:hist_rob_dec}
\end{figure}


\begin{figure*}
\includegraphics[scale=0.7]{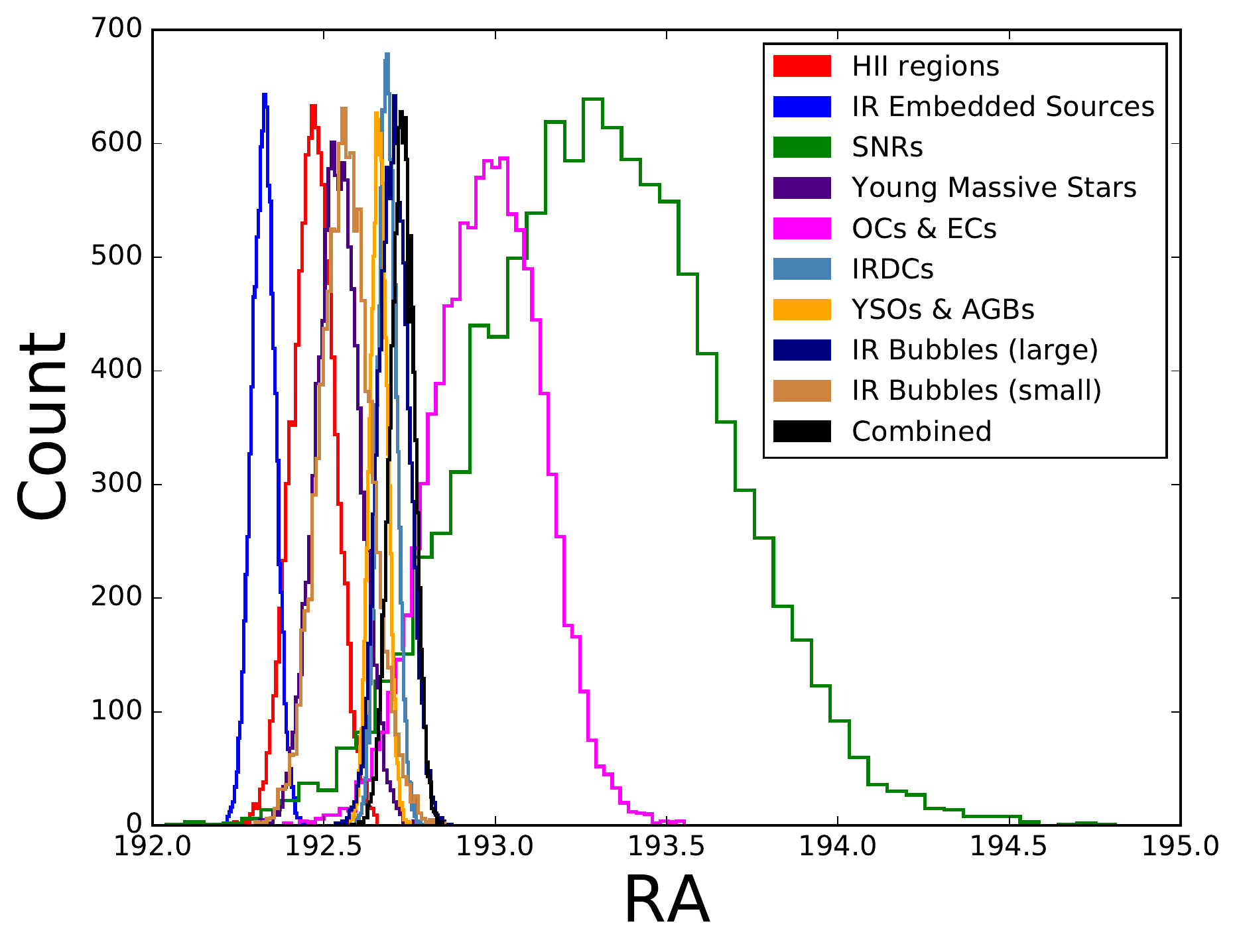}
\caption[]{Histograms of $\alpha$ for the estimated North
Galactic Pole for all of the samples analysed.}
\label{fig:hist_ra}
\end{figure*}


\begin{figure*}
\includegraphics[scale=0.7]{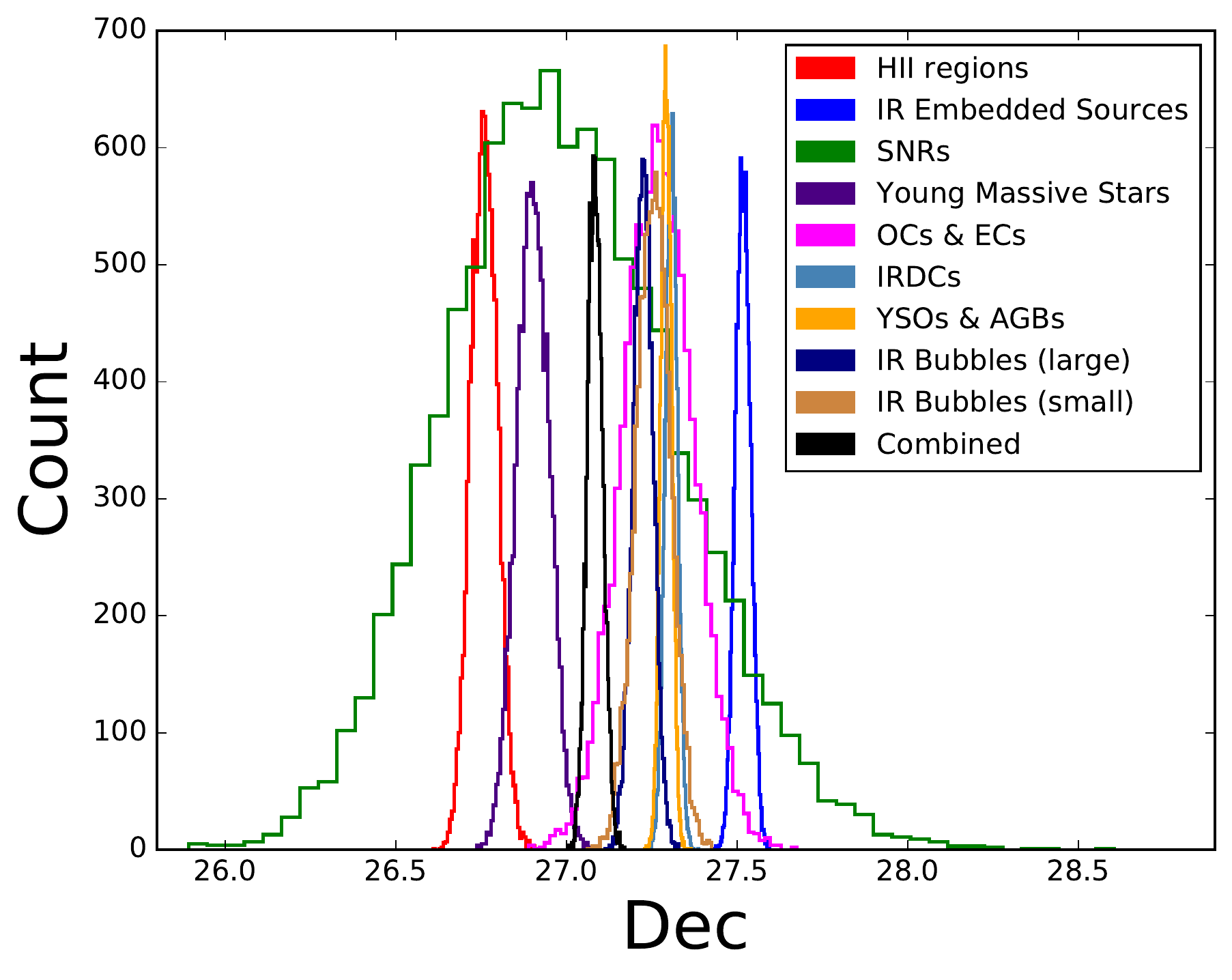}
\caption[]{Histograms of $\delta$ for the estimated North
Galactic Pole for all of the samples analysed.}
\label{fig:hist_dec}
\end{figure*}


\begin{table*}
\caption[]{Galactic Tracer Catalogues and Their Respective Estimated North Galactic Poles}
\centering
\begin{tabular}{l l r c c c}
\hline
Catalogue&Tracer Type&N$_{obj}$& $\alpha$ & $\delta$ & PA\\
&&&($^{\circ}$)&($^{\circ}$)&($^{\circ}$)\\
\hline
\protect\cite{Anderson14}  & \hii\, regions    & 8399     & 192.471$\,\pm\,$0.061 & 26.760$\,\pm\,$0.039 & 122.812$\,\pm\,$0.028\\
\hline
\protect\cite{Peretto09}   &IRDCs           &11303 & 192.684$\,\pm\,$0.028 & 27.307$\,\pm\,$0.018 & 122.908$\,\pm\,$0.013\\
\hline
\protect\cite{Robitaille08}&YSO \& AGB stars&18949 & 192.595$\,\pm\,$0.032 & 27.358$\,\pm\,$0.014 & 122.896$\,\pm\,$0.011\\
\hline
\protect\cite{Morales13}   & OCs \& ECs     &695   & 192.983$\,\pm\,$0.155 & 27.266$\,\pm\,$0.104 & 123.043$\,\pm\,$0.071\\
\hline
\protect\cite{Green14}     & SNRs           &294   & 193.308$\,\pm\,$0.353 & 26.992$\,\pm\,$0.335 & 123.193$\,\pm\,$0.161\\
\hline
\protect\cite{Simpson12}& IR Bubbles (Large)&3744  & 192.705$\,\pm\,$0.043 & 27.224$\,\pm\,$0.029 & 122.917$\,\pm\,$0.020\\
\hline
\protect\cite{Simpson12}& IR Bubbles (Small)&1362  & 192.563$\,\pm\,$0.076 & 27.255$\,\pm\,$0.052 & 122.852$\,\pm\,$0.034\\
\hline
\protect\textit{\cite{Simpson12}}& \textit{IR Bubbles (Combined)}&\textit{5106}  & \textit{192.673}$\,\pm\,$\textit{0.038} & \textit{27.230}$\,\pm\,$\textit{0.026} & \textit{122.903}$\,\pm\,$\textit{0.018}\\
\hline
\protect\cite{Csengeri14}&IR Embedded Sources&10861& 192.325$\,\pm\,$0.034 & 27.517$\,\pm\,$0.022 & 122.742$\,\pm\,$0.016\\
\hline
\protect\cite{Lumsden13}   &Young Massive Stars & 2811 & 192.542$\,\pm\,$0.064 & 26.903$\,\pm\,$0.049 & 122.844$\,\pm\,$0.029\\
\hline
{\bf Combined} &  & {\bf 58405}& {\bf 192.729$\,\pm\,$0.035} & {\bf 27.084$\,\pm\,$0.023} & {\bf 122.928$\,\pm\,$0.016}\\
\hline
\end{tabular}
\label{tab:Poles}
\end{table*}


\begin{figure*}
\includegraphics[scale=0.6]{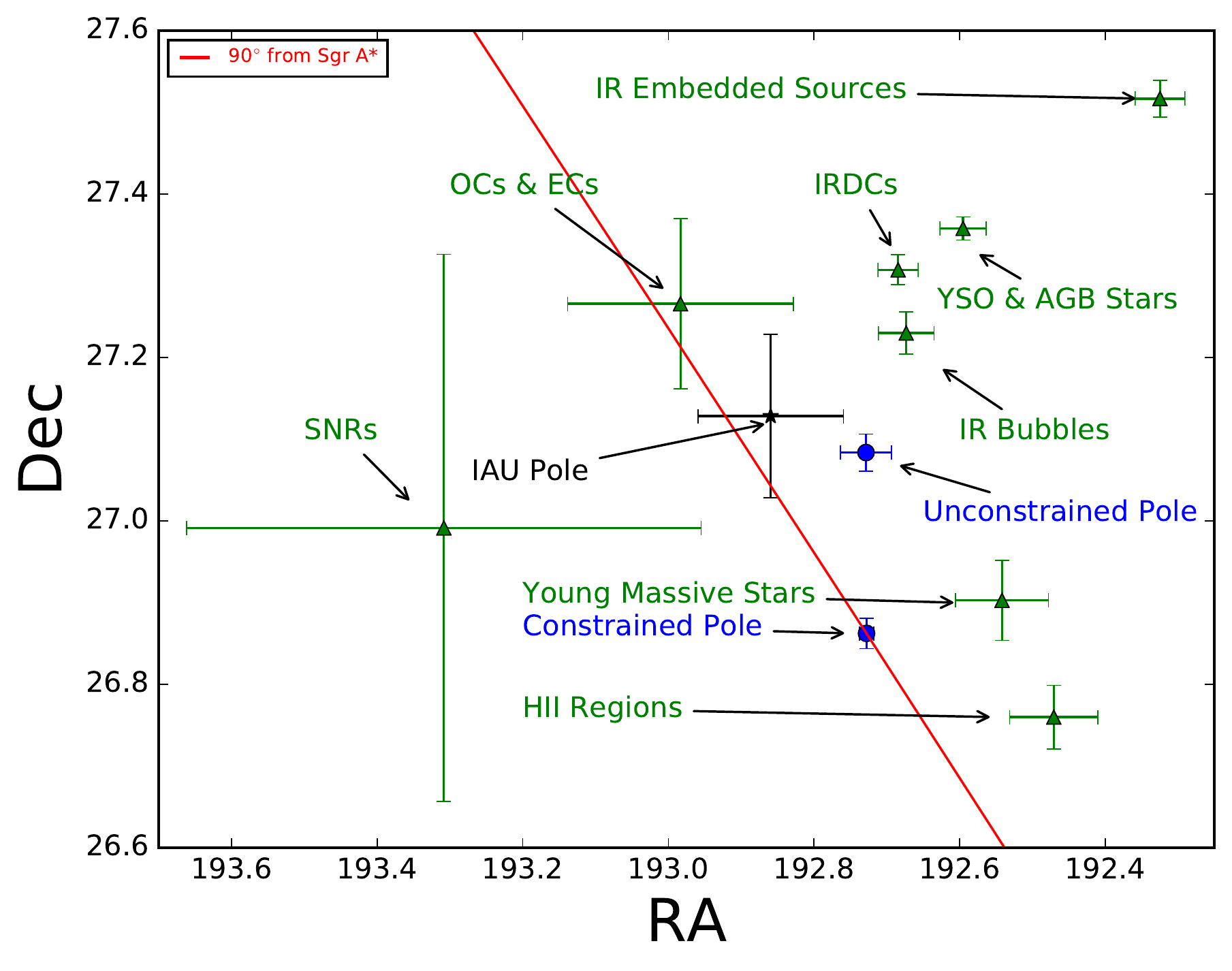}
\caption[]{Estimation of the North Galactic Pole. This Dec vs. RA plot
  shows the locations and the 1$\sigma$ uncertainties of the NGPs for
  each of the catalogues (shown in green triangles), the proposed constrained and unconstrained
  NGPs (shown in blue circle) and the current IAU NGP (shown in black
  star). A line (shown in red) is also shown tracing exactly
  90$^{\circ}$ from Sgr A*, using its position from
  \citet{Reid04}. Theoretically, the NGP should lie somewhere along
  this line.}
\label{fig:Poles_pic}
\end{figure*}

\subsection{Unconstrained NGP Solution}
\label{subsec:Method1}

In this method, we make the fundamental assumption that approximately
half of the Galactic tracer are above and half are below the Galactic
plane.
As discussed in Section~\ref{sec:Data}, we have specifically selected
catalogues for which this is a reasonable assumption.
Thus if we do a least-squares fit analysis and calculate the plane of
best-fitting of the tracers, this plane should coincide with the
Galactic plane, and the normal vector to this plane should point
towards the NGP.
For this analysis, we convert the ICRS positions of the tracers into
Cartesian coordinates on the equatorial system at unit distance.
Thus, the equation of the plane is given by,

\begin{center}
\begin{eqnarray}
\label{eq:galplane}
  z = Ax + By + C
\end{eqnarray}
\end{center}

\noindent where $A, B, C$ are the normalized coefficients.\\


The Cartesian coordinates of the tracers were then fitted to
Equation~\ref{eq:galplane} using the {\it linalg.lstsq} function from
the Python package \textit{Scipy} \citep{scipy} to obtain the values
of the normalized coefficients.
From these coefficients, the coordinates of the NGP on the ICRS were
calculated using the following relations,

\begin{center}
\begin{eqnarray}
& \alpha_{NGP} =  \arctan (B/A) &   \label{eq:ngp_alpha}\\
& \delta_{NGP} =  \arcsin (A^{2}+B^{2}+1)^{-1/2}   \label{eq:ngp_delta}
\end{eqnarray}
\end{center}


Because the tracers represent a tiny sample of the entire galactic
population, we used the bootstrap resampling method \citep{Efron79} to
obtain a statistically robust estimate of the coefficients.
Bootstrapping is the ideal resampling method to estimate the
properties of the estimators because it is distribution-independent,
provides a good estimation for small sample size, and is not affected
by outliers \citep{Ader08}.
By using the bootstrapping method, we generated 10$^4$ virtual
catalogues for each of the original catalogues.
We fitted these virtual catalogues to Equations~\ref{eq:galplane},\ref{eq:ngp_alpha}
and \ref{eq:ngp_delta} to obtain 10$^4$ values of
($\alpha_{NGP}$, $\delta_{NGP}$).\\


Lastly, histograms of these values were generated and the histograms
of the bootstrapped values appear to be Gaussian (e.g. see
Figure~\ref{fig:hist_rob_ra} and Figure~\ref{fig:hist_rob_dec}).
The means and the variances of these histograms are quoted as the
central values and uncertainties\footnote{All uncertainties for
  calculated quantities in this paper are 1$\sigma$.} respectively in
Table~\ref{tab:Poles}.
Combined histograms of RA and Dec of the NGPs determined from all the
catalogues are shown in Figure~\ref{fig:hist_ra} and
Figure~\ref{fig:hist_dec} respectively.
We used the ICRS coordinate of Sgr A* from \citet{Reid04} (listed in
\S1) to calculate the position angles.\\


The final ``unconstrained' NGP and position angle were obtained by
concatenating all the catalogues into one master catalogue and
applying the same method as described above.
These values are shown in bold in Table~\ref{tab:Poles}.
The NGPs for the individual tracer samples ({\it green triangles}) and
for all of the samples combined ({\it blue circle}) are plotted in
Figure~\ref{fig:Poles_pic}, along with the current (1958) IAU pole
({\it black star}).
The {\it red line} plotted Figure~\ref{fig:Poles_pic} is the arc that
lies exactly $90^{\circ}$ away from Sgr A*.
We discuss the significance of our best-fitting unconstrained NGP lying
more than 90$^{\circ}$ away from Sgr A* in \S4.2.


\subsection{Constrained NGP Solution \label{subsec:Method2}}

\begin{figure}
\includegraphics[scale=0.3]{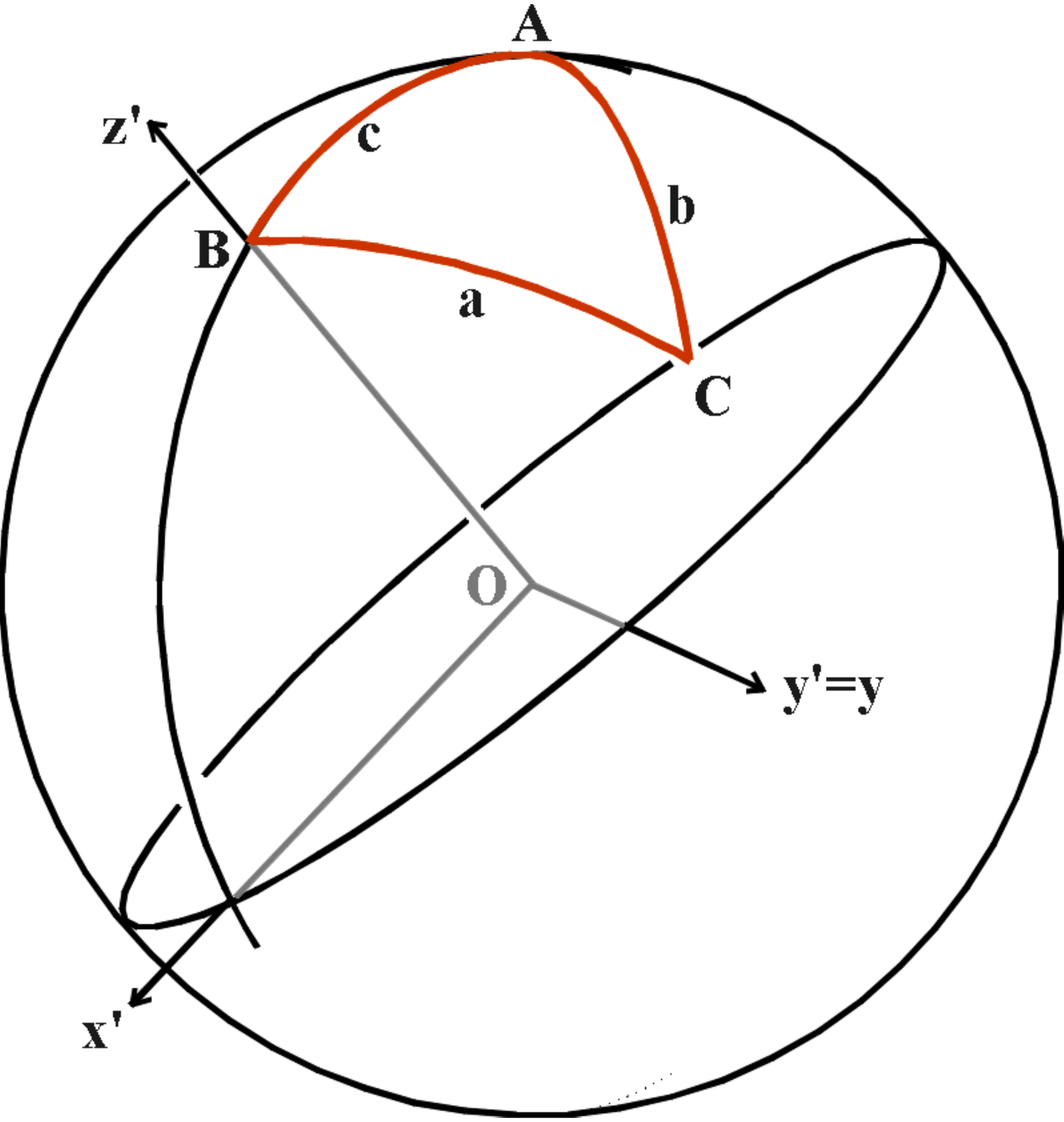}
\caption[]{A spherical triangle denoting the locations of 
the NCP (point A), NGP (point B), Sagittarius A* (point C).}
\label{fig:Method2}
\end{figure}

In this method we constrained the pole to be exactly $90^{\circ}$ away
from Sgr A*, along the great circle connecting Sgr A* with the
unconstrained NGP determined in \S3.1.
Consider a spherical triangle with the three vertices A,B and C (as
shown in Figure~\ref{fig:Method2})\footnote{Courtesy of
  http://star-www.st-and.ac.uk/}, where A corresponds to the North
Celestial Pole (NCP), B corresponds to the NGP and C corresponds to
Sgr A*.
In this triangle the known values are:

\begin{flalign*}
& BC = a = 90^{\circ} & \\
& AC = b = 90^{\circ} - \delta_{\rm Sgr A*} = 119^{\circ}.0078 & \\
& \angle ABC = B = \text{position angle} = 122^{\circ}.9280
\end{flalign*}

\noindent Since the two sides and the non-included angle is known,
unique solutions exist if $a > \arcsin{(\sin{b}\sin{A})}$, which is
the case for our problem.
Thus, we solved for $A$ and the side $c$ by using the spherical laws
of sines and Napier's analogies:

\begin{flalign*}
A &= \arcsin{\left(\frac{\sin{a}\sin{B}}{\sin{b}}\right)} = 73^{\circ}.6890 & \\
c &= 2\arctan{\left[\tan{\left(\frac{b-a}{2}\right)}\frac{\sin{\left(\frac{B + A}{2}\right)}}{\sin{\left(\frac{B - A}{2} \right)}}  \right]} = 63^{\circ}.1375
\end{flalign*}

\noindent From Figure~\ref{fig:Method2}, we used the following
relations to solve for the NGP:

\begin{flalign*}
& \alpha_{pole} = \alpha_{Sgr A*} - A = 192^{\circ}.7278 &\\
& \delta_{pole} = 90^{\circ} - c = 26^{\circ}.8625
\end{flalign*}

\noindent After propagating the uncertainties, we obtained the
following values for the ``constrained" NGP:

\begin{flalign}
& \alpha^{\text{constrained}}_{NGP} = 192^{\circ}.728 \,\pm\, 0^{\circ}.010 &\\	
& \delta^{\text{constrained}}_{NGP} = 26^{\circ}.863 \,\pm\, 0^{\circ}.019
\end{flalign}


\section{Discussion}
\label{sec:Discussion}

\subsection{Comparison with the IAU Pole}


The IAU North Galactic Pole was determined nearly six decades ago, and
heavily relied upon a single tracer (Galactic neutral hydrogen).
In our analysis, we have estimated the NGP using eight different
tracer samples, and the vast majority of the objects were unknown the
astronomers decades ago.
While most of the individual NGP determinations lie $>$2$\sigma$ away
from one another, our use of multiple tracer samples strengthens our
case that our best-fitting constrained and unconstrained NGP estimates are
not biased by any particular class of tracer (and is certainly not
overly dependent on a single tracer like the IAU pole).
Since the number of tracers used is a minuscule sample of the entire
galactic tracer population, we used the bootstrapping re-sampling
method to generate 10$^4$ synthetic catalogues for each tracer type.
Bootstrapping takes into account the potential distortion from poorly
representative samples.
Therefore, the number of tracers and the number of data points makes
our analysis more statistically robust when compared to the current
IAU definition of the NGP.\\


The current IAU pole is based on an estimate of the NGP with
uncertainty $\sim$0.$^{\circ}$1 \citep{Blaauw60}.
In comparison, our proposed poles (both constrained and unconstrained)
have an uncertainty of $\sim$0$^{\circ}$.03 in RA and
$\sim$0$^{\circ}$.02 in Dec, so we have attained a factor of
$\sim$4-5$\times$ improvement in accuracy for this important Galactic
parameter.


\subsection{Comparison with recent papers}


In the recent years, three studies have addressed the issue of
improving the Galactic coordinate system: \citet{Liu2011},
\citet{Liu11ir}, and \citet{Ding15}.
In this section, we compare our results with these papers in
chronological order.\\


\citet{Liu2011} noted that the GCS is rotating with respect to the
ICRS, and this rotation factor cannot be removed easily due to the
complexity of the fundamental catalogues, i.e. FK4 and FK5.
As a temporary solution, they derived the rotation matrix from the
equatorial to the GCS in the framework of the ICRS using the bias
matrix.
They also point out the necessity of placing Sgr A* at the centre of
the GCS and recommend that the IAU reconsiders the definition of the
GCS, which may result in a more accurate depiction of the GCS. In their 
proposed method, they utilize the ICRS coordinates of Sgr A* and determine
the coordinate of the NGP by setting the dot product of the ICRS coordinates
of Sgr A* and the ICRS coordinates of the NGP to be 0, i.e. by setting the
NGP exactly 90$^{\circ}$ away from Sgr A*. 
For example, using \citet{Reid04}'s estimation of the RA and Dec of Sgr*, 
their estimation of the fundamental parameters are -- $\alpha_{P}$ =
12$^{h}$51$^{m}$36$^{s}$.7151981, $\delta_{P}$ =
27$^{\circ}$06'11''.193172, $\theta$ = 123$^{\circ}$0075021536.
While their paper brought up the issue of the limitation of the current GCS for the 
first time in decades, it only used one observational data point
(\citet{Reid04}'s estimation of the coordinates of Sgr A*) to 
estimate the coordinates of the NGP, without considering 
additional data \citep{Liu11ir},thus lacking statistical rigour. 
Furthermore, in their following paper, \citet{Liu2011} 
acknowledged that this estimation is off by several arcminutes 
from their estimation that used the bias matrix.\\
%


In contrast, we used nine different tracers that were bootstrapped
10$^4$ times each, thus effectively our sample size consisted of
90,000 catalogues, making our estimates of the parameters and the
uncertainties statistically more robust.
We also used two methods to determine the pole, one which relies on
the location of Sgr A* and the other one does not. Thus, we show a
more complete analysis of parameter estimation.\\


\citet{Liu11ir} later used the 2MASS near-IR Point Source Catalogue
\citep{Cutri03, Skrutskie06} and SPECFINDv2.0, a radio cross-identification catalogue,
\citep{Vollmer10} in radio band to determine the fundamental
parameters.
They used two different methods to determine the fundamental
parameters for each catalogue type -- fixed z-axis, i.e. the z-axis of
the GCS points in the direction of the NGP exactly, and fixed x-axis,
i.e. the x-axis of the GCS points in the direction of the Sgr A*.
They divided the catalogues in 360 longitudinal bins and for each of
the fixed axis, they derived the parameters by determining the
equatorial positions of the geometrical centres of each
1$^{\circ}$-longitude bin. As a result, they published four possible sets
of fundamental parameters.
One of the catalogues, SPECFIND v2.0 is not evenly distributed and
shows bias towards the northern hemisphere
\citep{Liu11ir}. \citet{Liu11ir} acknowledged that the global
structure of the SPECFIND catalogue had a strong effect on their
results and they tried to remove biasing by restricting the data to
$\abs{b} \leq 5^{\circ}$. In addition, they ignored the weak
interstellar extinction at the infrared and radio bands.
Furthermore, as noted by \citet{Ding15}, \citet{Liu11ir} used only two
catalogues, whereas in order to trace the full physical feature of the
Milky Way, we need to consider more tracers. \citet{Ding15} also noted
that \citet{Liu11ir} did not offer any explicit recommendation for the
GCS and that the methods used to find the galactic plane requires
improvement.\\
%


\citet{Ding15} conducted a similar analysis building on the
methodology of \citet{Liu11ir}.
They used two all-sky surveys - the AKARI infrared all-sky survey
\citep{Murakami07} and the WISE all-sky catalogue \citep{Wright10} -
covering sources over six infrared bands between 3.4$\mu$m and 90
$\mu$m.
Their methods are similar to that of \citet{Liu11ir} -- fixed z-axis,
i.e. the z-axis of the GCS points in the direction of the NGP exactly,
and fixed x-axis, i.e. the x-axis of the GCS points in the direction
of the Sgr A*.
To obtain the galactic plane, they created 360 bins, each
corresponding to 1$^{\circ}$ galactic longitude and calculated the
medians of the bins, which was used to do the
least-squares-fitting.
The final results they proposed are: $\alpha_{P}$ = 192$^{\circ}$.777,
$\delta_{P}$ = 26$^{\circ}$.9298, $\theta$ = 122$^{\circ}$.95017,
based on the fixed x-axis method.
They argued that the x-axis method is a better option because in this
method they only calculated one parameter, the position angle, whereas
in the fixed z-axis method, they calculated two parameters and so it
maybe more prone to errors.
Furthermore, when quoting the final values, they averaged the values
obtained from the six wavelengths by applying equal weight to the
values.\\


We have identified two limitations regarding the methodology of \citet{Ding15}.
First, even though only one parameter, the position angle, was
calculated for the fixed x-axis method, \citet{Ding15} acknowledged
that their method was affected by singular points near the Galactic
Centre and the anti-Galactic Centre because the value of the position
angle became very unstable near these two longitudes, which resulted
in them discarding many values \citep[pages 7 \& 11 of ][]{Ding15}.
In comparison, we used bootstrapping resampling method to counter any
possible outlier effect.
As a result, we did not have to remove any data points from our analysis.
We also note that since the measurements of the position angle at
different wavelengths were affected differently and since there are different
number of data points in different bands, applying equal weight to
calculate the final value does not seem to be warranted. We further note that 
\citet{Ding15} did not include any uncertainty in
their measurement of the fundamental parameters.

\begin{table}
\caption[Estimates of the Position Angle Between Sgr A* and NGP]{Estimates of the Position Angle Between Sgr A* and NGP}
\centering
\begin{tabular}{l l l}
\hline
Study & PA    & unc.\\
...   & (deg) & (deg)\\
\hline
IAU \citep[calc. by ][]{Liu2011}$^1$ & 122$^{\circ}$.93192526   & ...\\
IAU \citep[calc. by ][]{Liu2011}$^1$ & 122$^{\circ}$.93191857   & ...\\
\citet[][Sgr A* fixed]{Liu2011}      & 123$^{\circ}$.0075021536 & ...\\
\citet[][{\it z}-fixed]{Ding15}      & 122$^{\circ}$.86216      & ...\\
\citet[][{\it x}-fixed]{Ding15}      & 122$^{\circ}$.95017      & ...\\
{\it This Study}                     & {\it 122$^{\circ}$.928}  & {\it 0$^{\circ}$.016}\\
\hline
\end{tabular}
$^1$ calculated by \citet{Liu2011} using different transformations
from FK4 (B1950.0) to FK5 (J2000.0) systems to the ICRS.
\label{tab:PA}
\end{table}


\subsection{Height of the Sun Above the Galactic Midplane}


Our unconstrained estimate of the NGP lies
90$^{\circ}.120\,\pm\,0^{\circ}.029$, rather than 90$^{\circ}$, away
from the dynamical centre of the Galaxy (Sgr A*).
As Sgr A* lies a finite distance away \citep[\ro\, = 8.2$\,\pm\,$0.1
  kpc;][]{Bland-Hawthorn16}\footnote{Amusingly, the most recent best
  estimate of the Galactocentric distance \ro\, advocated from the
  extensive literature survey of \citet{Bland-Hawthorn16} of
  8.2\,$\pm$\,0.1 kpc is identical in value to that adopted in the
  works that helped define the 1958 IAU GCS \citep{Blaauw60}.}, this
discrepancy can be explained by taking into account that the Sun has
some finite distance above the Galactic midplane \citep[i.e. the ratio
  of the Sun's height above the midplane compared to the distance to
  Sgr A* is very small, but not negligible or zero; see also \S3 and
  Fig. 3 of ][]{Goodman14}.
Hence there should be little surprise that the estimated NGP {\it
  should} be $>$90$^{\circ}$ from Sgr A*.\\

\begin{figure}
\includegraphics[scale=0.6]{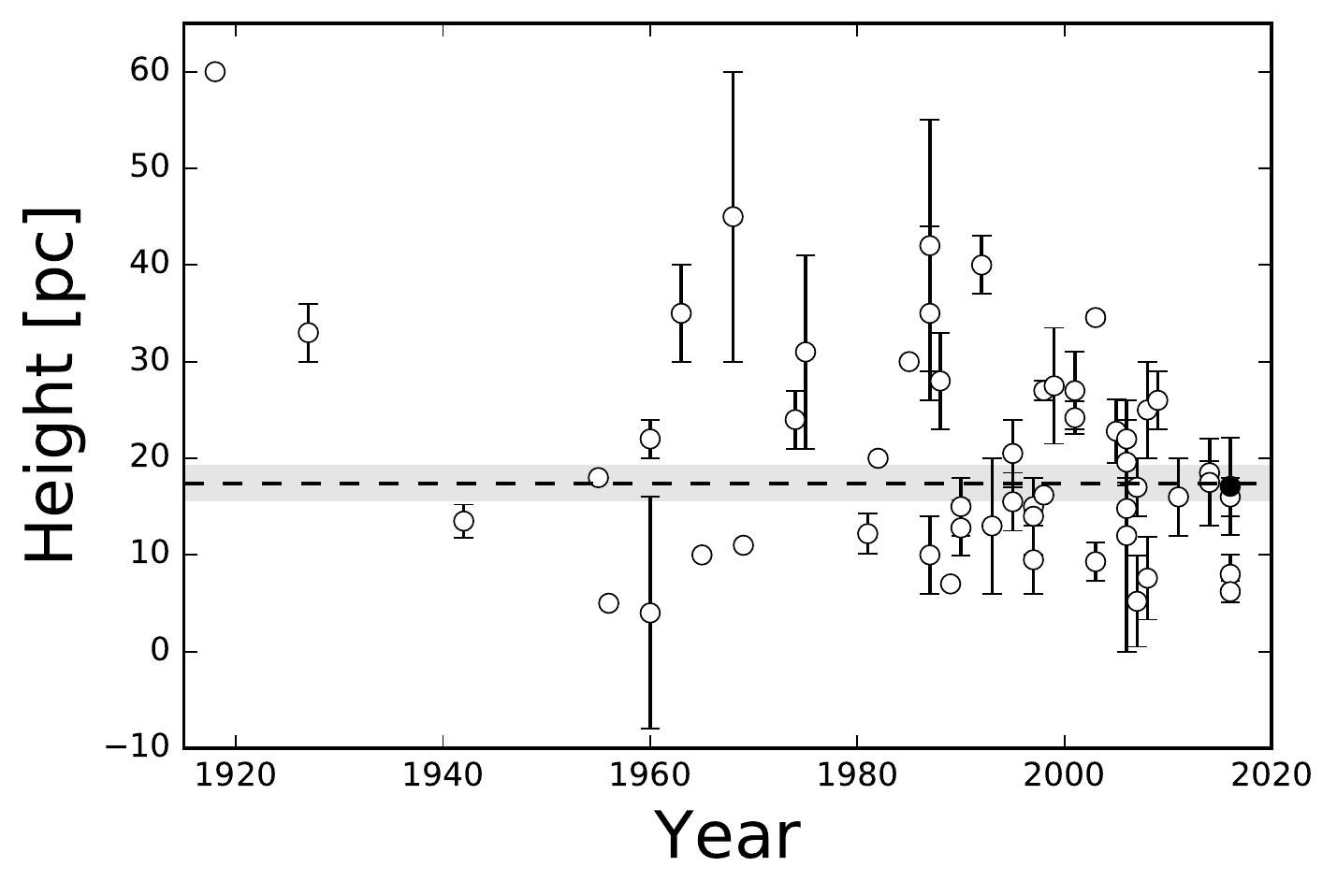}
\caption[]{Plot of publication year vs. estimated height of the Sun
  above the Galactic plane $z_{\odot}$ (in parsecs) calculated by
  various studies since 1918, except \citet{Kreiken26}'s estimate of 250 pc. 
  The {\it filled dot} represents our estimate, 
  the open dots represent values determined by different studies,
  the dashed line represents the median (17.4 pc) and the grey-shaded area the
  uncertainty range ($\pm$ 1.9 pc) of all  the measurements. 
  The published estimates are listed in Table \ref{tab:height}.}
\label{fig:height}
\end{figure}

One can geometrically estimate the height of the Sun above the
Galactic midplane using the estimate of the Galactocentric distance
(to Sgr A*) and the angular separation between the unconstrained
estimate of the Galactic plane and Sgr A*.
The geometry of this case is well illustrated in Fig. 3 of
\citet{Goodman14}, and not reproduced here.
The only difference with Fig. 3 of \citet{Goodman14} is that we
replace the ``{\it IAU mid-plane b=0}'' with our unconstrained
estimate of the Galactic plane using our combination of tracer samples
(\S3.1).
Combining the galactocentric distance from \citet{Bland-Hawthorn16}
with our estimate of the angular separation of our proposed NGP from
Sgr A*, we estimate that the Sun would need to be 17.1\,$\pm$\,5.0 pc
above the Galactic midplane in order to explain Sgr A*'s angular
distance below our best-fitting unconstrained Galactic Plane.
This value is only 1.2$\sigma$ below the value adopted by
\citet{Bland-Hawthorn16} in their recent review on the Milky Way
\citep[25\,$\pm$\,5 pc; based on the solar offset from the local disc
  midplane based on the SDSS photometric survey][]{Juric08}
In Table \ref{tab:height}, we compile a list of 55 other published
estimates of the Sun's height $z_{\odot}$ above the Galactic midplane,
and plot the values in Figure \ref{fig:height}.
Combined with our new estimate, the ensemble of 56 estimates has a
true median \citep{Gott01} of $z_{\odot}$ $\simeq$ 17.4\,$\pm$\,1.9
pc, with a 95\%\, confidence range of 15-22 pc, suggesting that our
estimate is very close to the locus of previous estimates.
Given the Sun's vertical velocity with respect to the Local Standard
of Rest \citep[7.25\,$\pm$\,0.36 \kms;][]{Schonrich10}, this suggests
that the Sun passed through the Galactic midplane approximately
2.5\,$\pm$\,0.3 Myr ago, coincidently at the beginning of the
Pleistocene epoch \citep[2.58 Myr ago;][]{Cohen15}.



\begin{table*}
  \caption[A Century of Estimates of the Sun's Height Above the Galactic Mid-Plane]{
    A Century of Estimates of the Sun's Height Above the Galactic Mid-Plane}
\begin{tabular}{l l c}
\hline
Reference & Description & $z_{\odot}$(pc)\\
\hline
\citet{Shapley18}       & Globular clusters & 60\\
\citet{Kreiken26}       & Local stellar system & 250\\ 
\citet{Gerasimovic27}   & Cepheids, O-, B-, c- \& ac-type stars  & 33$\,\pm\,$3\\   
\citet{vanTulder42}     & Cepheids, planetary neb., c-, O-, B-, and WR-type stars & 13.5$\,\pm\,$1.7\\
\citet{vanRhijn55}      & A-type stars  & 18\\
\citet{vanRhijn56}      & K-type stars  & 5\\
\citet{Gum60}           & \hi\, gas & 4$\,\pm\,$12\\
\citet{Blaauw60B}       & OB-type stars \& Cepheids & 22$\,\pm\,$2\\
\citet{Kraft63}         & Cepheids & 35$\,\pm\,$5 (30-40)\\
\citet{Elvius65}        & AFGK-type stars & 10\\
\citet{Fernie68}        & Cepheids & 45$\,\pm\,$15\\
\citet{deVaucouleurs69} & Galactic absorbing layer & 11\\
\citet{Stothers74}      & OB-type stars & 24$\,\pm\,$3\\
\citet{Stenholm75}      & WR stars & 31$\,\pm\,$10\\
\citet{Toller81}        & Pioneer 10 observation of background starlight & 12.2$\,\pm\,$2.1\\
\citet{Lynga82}         & Open clusters & 20\\
\citet{Magnani85}       & Molecular gas & 30\\
\citet{Stobie87}        & $UBV$ star counts & 42$\,\pm\,$13\\
\citet{Caldwell87}      & Cepheids & 35$\,\pm\,$9$^{a}$\\
\citet{Pandey87}        & Interstellar matter & 10$\,\pm\,$4\\
\citet{Pandey88}        & Open clusters & 28$\,\pm\,$5\\
\citet{Ratnatunga89}    & Stars in Yale Bright Star Catalogue & 7\\
\citet{Conti90}         & Wolf-Rayet stars & 15$\,\pm\,$3\\
\citet{Toller90}        & Background light \& interstellar dust & 12.8$\,\pm\,$2.9\\
\citet{Yamagata92}      & $UBV$ star-count & 40$\,\pm\,$3\\
\citet{Brand93}         & Local molecular clouds & 13$\,\pm\,$7\\
\citet{Hammersley95}    & Two Micron Galactic Survey, IRAS and COBE & 15.5$\,\pm\,$3\\
\citet{Cohen95}         & IRAS point source \& FAUST catalogues & 15.5$\,\pm\,$0.5\\
\citet{Humphreys95}     & Optical star counts & 20.5$\,\pm\,$3.5\\
\citet{Ng97}            & Star counts \& colors & 15\\
\citet{Binney97}        & COBE all-sky survey & 14$\,\pm\,$4\\
\citet{Reed97}          & OB-type stars & 9.5$\,\pm$\,3.5 (6-13)\\
\citet{Freudenreich98}  & Diffuse infrared background & 16.2$^a$\\
\citet{Mendez98}        & Interstellar dust & 27$\,\pm\,$1\\
\citet{Chen99}          & COBE \& IRAS all-sky reddening map & 27.5$\,\pm\,$6.0\\
\citet{Maiz-Apellaniz01}& OB-type stars & 24.2$\,\pm\,$1.7\\
\citet{Chen01}          & SDSS star counts & 27$\,\pm\,$4\\
\citet{Branham03}       & Hipparcos stars except OB stars & 34.56$\,\pm\,$0.56\\
\citet{Paladini03}      & \hii\, regions & 9.3$\,\pm\,$2 (7.3-11.3)\\
\citet{Joshi05}         & Reddening of open clusters & 22.8$\,\pm\,$3.3\\
\citet{Reed06}          & OB-type stars & 19.6$\,\pm\,$2.1\\
\citet{Piskunov06}      & Open clusters & 22$\,\pm\,$4\\
\citet{Elias06}         & OB-type stars & 12$\,\pm\,$12\\
\citet{Bonatto06}       & Open clusters & 14.8$\,\pm\,$2.4\\
\citet{vanLeeuwen07}    & Hipparcos A-/F-type stars & 5.2$\,\pm\,$4.7\\
\citet{Joshi07}         & Young open clusters \& OB-type stars & 17$\,\pm\,$3\\
\citet{Kong08}          & OB-type \& horizontal branch stars  & 7.6$\,\pm\,$4.3$^b$\\
\citet{Juric08}         & SDSS stellar density distribution & 25$\,\pm\,$5\\
\citet{Majaess09}       & Cepheids & 26$\,\pm\,$3\\
\citet{Liu11_height}    & Open clusters & 16$\,\pm\,$4\\
\citet{Buckner14}       & Open clusters & 18.5$\,\pm\,$1.2\\
\citet{Olausen14}       & Magnetar and magnetar candidates & 17.5\,$\pm$\,4.5 (13-22)\\
\citet{Bobylev16}       & OB-, Wolf-Rayet-type stars, Cepheids \& \hii\, regions & 16$\,\pm\,$2\\
\citet{Bobylev16B}      & \hii\, regions, masers \& molecular clouds  & 8$\,\pm\,$2 (6-10)\\
\citet{Joshi16}         & Open clusters & 6.2\,$\pm$\,1.1\\
{\bf This study}        & Sgr A* offset from Galactic plane & {\bf 17.1$\,\pm\,$5.0}\\
\hline
{\bf True Median ${z}_{\odot}$} & & {\bf 17.4\,$\pm$\,1.9}\\
\hline
\end{tabular}
\vspace{1pt}
\begin{flushleft}
Notes:\\
$^a$: Estimated value for \ro\,=\,8.2 kpc.\\
$^b$: Weighted mean using their two estimates (3.5$\,\pm\,$5.4 pc, 15.2$\,\pm\,$7.3 pc).
\end{flushleft}
\label{tab:height}
\end{table*}
\section{Conclusions}

Given the importance of the Galactic Coordinate System in the study of
galactic structure and kinematics, a more accurate representation of
the GCS based on the physical markers of the Galaxy may be overdue.
We have derived the fundamental parameters that define the Galactic
Coordinate System, i.e. the RA and Dec of the NGP and the position
angle of the NGP with respect to the Galactic Centre at the North
Celestial Pole, using two different analyses on large samples of
Galactic tracers.
In the unconstrained method, we used eight galactic tracers to
determine the plane of best-fitting to determine the location of the
NGP, without taking the location of Sgr A* into account. The
parameters determined using this method are: $\alpha_{NGP}$ =
$192^{\circ}.729$ $\,\pm\,$ $0^{\circ}.035$,
$\delta_{NGP}$ = $27^{\circ}.084$ $\,\pm\,$ $0^{\circ}.023$ and 
$\theta\,$ = $122^{\circ}.928$ $\,\pm\,$ $0^{\circ}.016$.

Using this first method, the NGP lies
$90^{\circ}.120\,\pm\,0^{\circ}.029$ away from Sgr A*, instead of
being exactly $90^{\circ}$ away.
We explain this discrepancy as a result of the height of the Sun above
the Galactic midplane.
Using this discrepancy, we have independently estimated the the height
of the Sun above the Galactic midplane to be $z_{\odot}$ =
$17.1\,\pm\,5.0$.
Our new estimate of the height of the Sun above the Galactic midplane
is similar to the true median estimate based on 56 published estimates
(17.4\,$\pm$\,1.9 pc).

Using the ``constrained'' methodology, we solved for an estimate of
the NGP which would lie {\it exactly} $90^{\circ}$ away from Sgr A*,
along the great circle connecting Sgr A* with the unconstrained
NGP.
The parameters determined using this method are: $\alpha_{NGP}$ =
$192^{\circ}.7278 \,\pm\, 0^{\circ}.0098$,
$\delta_{NGP}$ = $26^{\circ}.863 \,\pm\, 0^{\circ}.019$ and 
$\theta\,$ = $122^{\circ}.928$ $\,\pm\,$ $0^{\circ}.016$.
In both the methods, the uncertainties in the position of the NGP are
$\sim$4-5$\times$ smaller than those of the original IAU estimate from
the 1950s.

Which solution is the ``best'' may depend on the application.
For a simple 2D Galactic coordinate system, the constrained (2nd)
solution which forces the poles to be exactly 90$^{\circ}$ from Sgr A*
may be a more reasonable option as it anchors the origin to the
dynamical centre of the Galaxy.
However, for Galactic kinematic calculations, one would want to solve
for the motion of the stars (or Sun) in a Galactocentric cylindrical
coordinate system, preferably one which takes into account the Sun's
height above the Galactic midplane ($z_{\odot}$ = 17 pc).
In this case, the parameters for the unconstrained NGP should be more
the more relevant solution.
As the Sun lies a finite distance above the Galactic midplane, one
should not expect the dynamical centre of the Galaxy (Sgr A*) to
appear precisely 90$^{\circ}$ from the normal to the best-fitting Galactic
plane determined using the positions of Galactic tracers as seen from
the solar system.
In this case, one would expect Galactic tracers to define a Galactic
plane projected asymptotically to lie some small angle {\it above} Sgr
A*, as is seen.
With the recent release of the first data release of {\it Gaia}
astrometry, it will be interesting to compare geometric estimates of
the NGP based on Galactic tracers with dynamical estimates based on
stellar kinematics.

\section*{Acknowledgements}

MTK participated in the summer 2014 University of Rochester REU
program supported by PHY-1156339.
MTK and EEM acknowledges support from NSF awards AST-1313029.
EEM acknowledges support from the NASA NExSS program. 
We thank Eric Feigelson, Segev BenZvi, Sanha Cheong, and Alec Kirkley
for discussions on statistics, and Matthias Steinmetz for discussions
at the 2015 IAU meeting on the Milky Way.

\bibliographystyle{mn3e}
\bibliography{mamajek}


\label{lastpage}

\end{document}